\documentclass{ceab}   
\usepackage{epsfig}     
\usepackage{graphicx}   
\usepackage{hyperref}
\usepackage{ceabbib}     
\usepackage[T1]{fontenc}
\usepackage{amsmath, amsthm, amsfonts}
\usepackage{textcomp}
\usepackage{subfig}
\usepackage{rotating}
\usepackage{setspace}

\def\gore{CME volume calculation from 3D GCS reconstruction}

\begin{document}

\title{CME volume calculation from 3D GCS reconstruction}

\author{L.~Holzknecht$^{1}$, M.~Temmer$^{1}$, M.~Dumbovi\'c$^{1}$, S.~Wellenzohn$^{1}$\\K.~Krikova$^{1}$, S.G.Heinemann$^{1}$, M.~Rodari$^{1,2}$, B.~Vr\v{s}nak$^3$ and A.M.~Veronig$^{1}$ 
\vspace{2mm}\\
\it $^1$IGAM/Institute of Physics, University of Graz, Austria\\ 
\it $^2$Institute of Physics G. Occhialini, University of Milano-Bicocca, Italy\\
\it $^3$Hvar Observatory, Faculty of Geodesy, University of Zagreb\\
Correspondance: \href{mailto:lukas.holzknecht@edu.uni-graz.at}{lukas.holzknecht@edu.uni-graz.at},    \href{mailto:manuela.temmer@uni-graz.at}{manuela.temmer@uni-graz.at}
}
 
\maketitle

\begin{abstract}
The mass evolution of a coronal mass ejection (CME) is an important parameter characterizing the drag force acting on a CME as it propagates through interplanetary space. Spacecraft measure in-situ plasma densities of CMEs during crossing events, but for investigating the mass evolution, we also need to know the CME geometry, more specific, its volume. Having derived the CME volume and mass from remote sensing data using 3D reconstructed CME geometry, we can calculate the CME density and compare it with in-situ proton density measurements near Earth. From that we may draw important conclusions on a possible mass increase as the CME interacts with the ambient solar wind in the heliosphere. In this paper we will describe in detail the method for deriving the CME volume using the graduated cylindrical shell (GCS) model \citep[][see \figurename~\ref{fig:GCSModel}]{thernisien06,thernisien09}. We show that, assuming self-similar expansion, one can derive the volume of the CME from two GCS parameters and that it furthermore can be expressed as a function of distance.
\end{abstract}


\section{Introduction}
The mass evolution of a coronal mass ejection (CME) can be studied in the low to mid corona using white-light coronagraph data. These observations lead to the conclusion that the observed mass increase is most probably caused by an ongoing outflow from the Sun feeding plasma into the rear part of the CME structure \citep[see][]{bein13, howard18}, as well as provided mass from the erupting flux rope by magnetic reconnection in the current sheet beneath. As the CME reaches the Alfv\'enic critical point beyond about 20 solar radii (R$_{\odot}$), it can be assumed that an outflow from the solar surface is not able to account for a CME mass increase \citep{hundhausen1972}. In turn, with the expansion of the CME in interplanetary space, a pile-up at the CME front might become important \citep[see e.g.,][]{lugaz05,foullon07}. Added mass has important effects on the propagating CME. A strong pile-up would increase the drag which affects the CME travel time, impact speed as well as the energy input into the magnetosphere. For studying mass increase, we may calculate the CME density close to the Sun and compare it with in-situ proton density measurements near Earth. This study is a preparatory work for investigating mass pile-up in interplanetary space due to the interaction of the CME with the ambient solar wind. We describe in detail the method to derive the CME volume from GCS reconstruction that can be used to estimate the CME density close to the Sun from remote sensing data. 

\section{Method}
Using white-light coronagraphic images from the \textit{Sun Earth Connection Coronal and Heliospheric Investigation} \citep[SECCHI;][]{howard08} onboard \textit{Solar Terrestrial Relations Observatory} (STEREO) and the \textit{Large Angle and Spectrometric Coronagraph} \citep[LASCO;][]{Brueckner95} onboard the \textit{Solar and Heliospheric Observatory satellite} (SOHO) the 3D geometry of a CME can be obtained. Working under the assumption of self-similar expansion the graduated cylindrical shell (GCS) model can be used to derive the 3D geometry of a CME up to about 15 to 20 R$_{Sun}$ \citep{2011ApJS..194...33T}. The GCS model represents the idealized flux rope structure of a CME and is composed of two cones describing the legs of a CME and a torus-like structure connecting these cones. Therefore it is also known as the "hollow croissant". The equations that define the geometry of the model reflect both the self-similar expansion \citep{1538-4357-490-2-L191,2004AGUFMSH22A..04S} and the 3D morphology of the flux rope \citep{1985JGR....90.8173H}. These equations form the basis for calculating the total volume of the CME structure as defined by GCS reconstruction.   

\begin{figure}[htb]
	\includegraphics[scale=0.5]{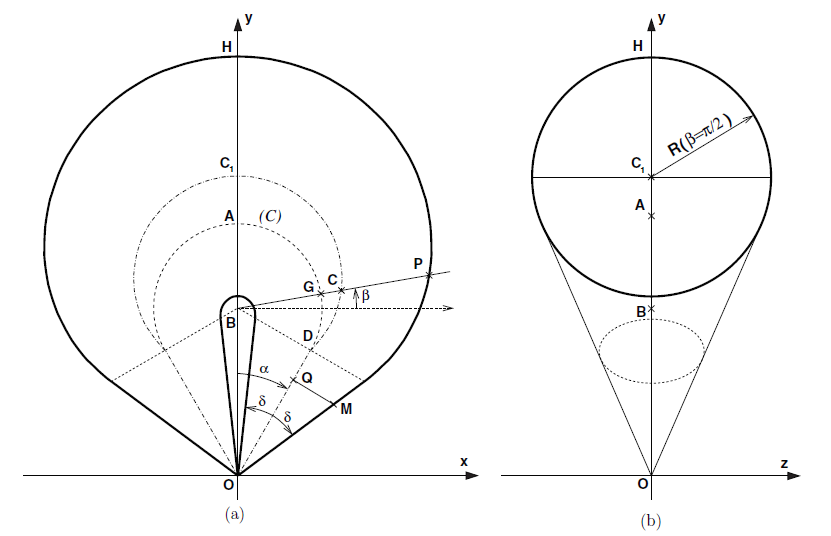}
	\caption{Schematic of the GCS model. The left panel shows a (O, x, y) planar cut of the croissant viewed face-on. The z-axis points toward the reader. The right panel shows a cut in the (O, y, z) plane where the croissant is viewed edge-on. In this view, only the circle (solid line) is in the (O, y, z) plane \citep[Figure 1 in][]{2011ApJS..194...33T}.}
	\label{fig:GCSModel}
\end{figure}

\begin{figure}[htb]
    \includegraphics[scale=0.45]{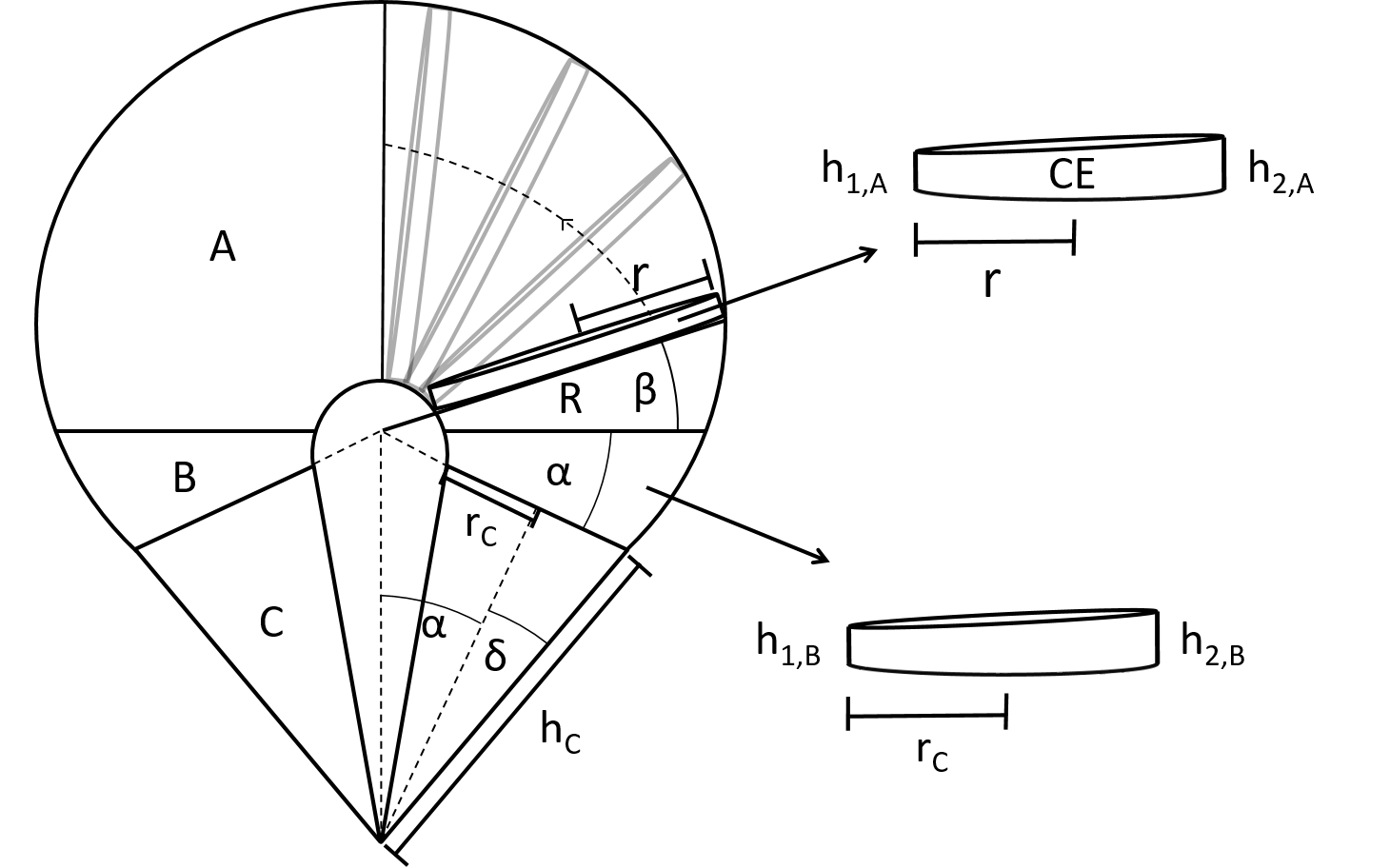}
	\caption{Sketch for calculating the CME volume from the GCS reconstructed flux rope showing the different sectors of the CME front part (A, B) and CME legs (C) as well as the parameters used in Equations 1--9.}\label{fig:GCSVolume}
\end{figure}

In order to approximate the volume of the entire GCS structure, we separate it into three different parts covering the frontal part of the CME and its legs (A, B, C; see \figurename~\ref{fig:GCSVolume}). This separation may also be used to estimate the density distribution within different parts of the CME separately. 

For Volume A we extract cylindrical elements (CE). We cut these elements for $\beta$ = 0\textdegree{} to 90\textdegree{} in small fractions of $\beta$ and call these fractions $\gamma$ (constant for all CE). Note that each of these elements has two different heights. These heights can be calculated by the distance R and the cross section radius $r$ as given by equations (19) and (27) in \cite{2011ApJS..194...33T}. Calculating the volume of these elements and summing them up comprises volume A. Volume B is calculated by a similar approach, namely a CE that is cut between the upper end of the cone and the x-axis (corresponding to $\beta$ = 0). Volume C simply represents the volume of a cone.

As the structure is axisymmetric, the total volume V$_{Total}$ of the GCS model is two times the sum of A, B, and C. For the entire volume calculation only two GCS parameters, the half-angle $\alpha$ and the half-angle of the cone $\delta$, are required. As we assume that the CME propagates in interplanetary space in a self-similar manner, these values are  constant over distance. 

\setlength\abovedisplayshortskip{0pt}
\setlength\belowdisplayshortskip{0pt}

\begin{equation}
    h_{\rm 1,A} = \tan(\gamma) R - 2r
\end{equation}

\begin{equation}
    h_{\rm 2,A}=\tan(\gamma)R
\end{equation}

\begin{equation}
    V_{\rm CE}=r^2\pi\frac{h_1+h_2}{2}
\end{equation}

\begin{equation}
    V_{\rm A}=\sum V_{\rm CE}
\end{equation}

\begin{equation}
    h_{\rm 1,B}=\sin(\alpha)(R-2r[\beta=0])
\end{equation}

\begin{equation}
    h_{\rm 2,B}=\sin(\alpha)R[\beta=0]
\end{equation}

\begin{equation}
    V_{\rm B}=r_c^2\pi\frac{h_{1,B}+h_{2,B}}{2}
\end{equation}

\begin{equation}
    V_{\rm C}=r_c^2\pi\frac{h_C}{3}
\end{equation}

\begin{equation}
    V_{\rm Total}=2(V_A+V_B+V_C)
\end{equation}

\section{Results}
To estimate the range of CME volumes, we use values for $\alpha$ of 5\textdegree{} to 70\textdegree{} and for $\kappa$ of 0.1 to 0.9 as given in the HELCATS catalogue covering 122 CMEs that were reconstructed with GCS (\url{https://www.helcats-fp7.eu/catalogues/wp3_kincat.html}). The CMEs occurred between the deep minimum in 2007 until the maximum phase in late 2013. The study of \cite{2004JGRA..109.7105Y} which consists of over 700 CMEs covering solar cycle 23, shows a similar range. 

\begin{figure}[htb]
	\includegraphics[width=\columnwidth]{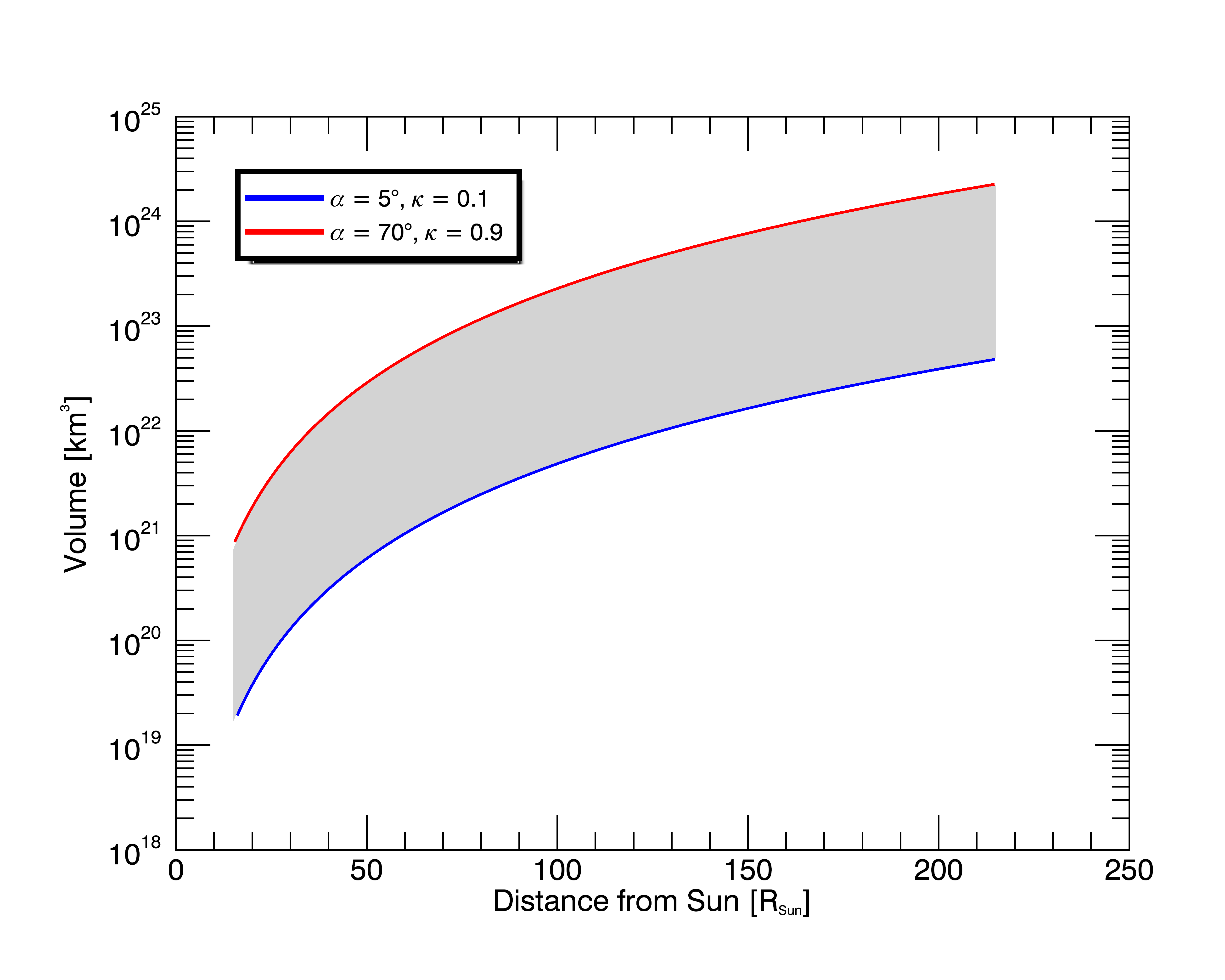}
	\caption{Results for the GCS volume with different input parameters. The grey shaded area displays the possible results for physically meaningful input parameters. $\alpha$ represents the half-angle and $\kappa$ is the sine of $\delta$ which represents the half-angle of the cones (see \figurename~\ref{fig:GCSModel} and \ref{fig:GCSVolume}).} \label{fig:volume_propagation}
\end{figure}

From these statistical input parameters we calculate a range of CMEs volumes as a function of distance from the Sun (see \figurename~\ref{fig:volume_propagation}). The minimum values range from approximately $10^{19}$ km$^3$ at 15 R$_{Sun}$ to almost $10^{23}$ km$^3$ at 215 R$_{Sun}$, whereas the largest CMEs have volumes of $10^{21}$ km$^3$ at 15 R$_{Sun}$ to some $10^{24}$ km$^3$ at 1 AU.

\section{Summary and conclusion}
In this paper we present a method to calculate CME volumes using the 3D geometry of a CME from GCS reconstruction. To derive the CME volume, two parameters are needed from GCS, namely the angle of the CME half-width ($\alpha$) and the angle of the half-width of the CME legs ($\kappa$). Assuming self-similar expansion, the derived CME volume can be given as a function of distance. Using typical values for $\alpha$ and $\kappa$, found in literature and recent catalogues, we obtain that the volume for small and large CMEs at a distance of Earth is spanning approximately two orders of magnitude.

The density is the only parameter to compare with in-situ measurements at different spacecraft locations and, hence, to study a possible mass pile-up of CMEs in interplanetary space. Since density represents the fraction of mass over volume, we also need to know the mass of the CMEs. We derive it from white-light remote sensing data \cite[preferably the 3D or total mass of a CME which can be estimated from stereoscopic white-light data as e.g., given by][]{bein13}. Considering typical values from $10^{14}$ g to $10^{17}$ g we have a distribution over three orders of magnitude \citep{bein13,vourlidas00}. 

Assuming that the mass stays constant beyond 20~R$_{Sun}$ enables the calculation of the CME density at specific locations. Comparing the value ranges of mass and volume for 1 AU, in a first order approach we estimate a density variation over about one order of magnitude, which is comparable to actual measurements \citep[see e.g.,][]{venzmer18}. 

This study, enables the derivation of the volume and will serve as a basis for studying variations of mass due to the interaction of CMEs with the ambient solar wind.

\section*{Acknowledgements} 
 M.T. and S.G.H acknowledge the support by the FFG/ASAP Program under grant no. 859729 (SWAMI). B.V. acknowledges the support by Croatian Science Foundation under the project no. 7549 "Millimeter and submillimeter observations of the solar chromosphere with ALMA" (MSOC). M.D. acknowledges funding from the EU H2020 MSCA grant agreement No 745782 (project ForbMod). The HELCATS catalogue builds on the work undertaken during the EU FP7 AFFECTS project. The catalogue provides analysis of CME properties based on the COR2 coronagraph observations and the Graduated Cylindrical Shell model. This is a pre-release version in the context of HELCATS since the events have not yet been correlated with the released WP2 catalogue. 
\bibliographystyle{ceab}


\end{document}